\documentstyle[aps,preprint,psfig]{revtex}

\begin{document}
\title
{\bf Dissipative phenomena in  chemically non-equilibrated quark gluon plasma}
\author
{D. Dutta, A. K. Mohanty, K. Kumar and R. K. Choudhury}
\address{ Nuclear Physics Division, Bhabha Atomic Research Centre,
Mumbai-400085, India}
\date{\today}
\maketitle
\begin{abstract}
The dissipative corrections to the hydrodynamic equations describing the
evolution of energy-momentum tensor and parton densities are derived in
a simple way using the scaling approximation for the expanding quark gluon
plasma at finite baryon density. This procedure has been extended to study the
process of chemical equilibration using a set of rate equations appropriate
for a viscous quark gluon plasma. It is found that in the presence of dissipation, the
temperature of the plasma evolves slower, whereas the quark and gluon
fugacities evolve faster than their counterparts in the ideal
case without viscosity.
\end{abstract}
PACS number(s): 25.75.Dw,12.38Mh,12.38.Bx,24.85.+p
\section{INTRODUCTION}
A phase transition from normal nuclear matter to a quark gluon plasma
(QGP) is expected in ultrarelativistic heavy ion collisions planned to be carried at
the future colliders of RHIC and LHC. If the QGP is formed during the collision,
according to the usual scenario, it is thought that a local thermal equilibrium
will set in over a typical time scale of 1 fm/c \cite{1}. The plasma would then
expand hydrodynamically and cool down until it reaches a temperature
$T_c$ when the confinement phase transition starts.  However, some recent
studies \cite{2,3} have suggested that the plasma may attain kinetic equilibrium in a
very short time scale of $\tau=0.2-0.7$ fm/c, but it may not be chemically equilibrated.
Earlier theoretical studies \cite{4,5,6} have revealed that a chemically non-equlibrated
plasma cools faster than the equilibrated plasma, when the plasma is assumed to
be ideal i.e. non dissipative. But in a realistic scenario  the dissipative
phenomena, in principle, should play an important role \cite{7}.
 It needs to be mentioned here that the dissipative effects in fluid
hydrodynamics can be neglected when the
 Reynold's number ($\approx L/\lambda$, L is the dimension of the medium and $\lambda$
 is the mean free path of the constituents) is quite large. However, in
 nucleus-nucleus collisions, as the dimension of the plasma is not large compared
 to the mean free path of the partons, dissipative effects are expected to be
important during the hydrodynamic expansion of the plasma.
The formalism of dissipative hydrodynamics is, in general, complicated since the
form of dissipative corrections depends on the choice of the rest frame \cite{7,8,9,10}.
The effects of dissipative phenomena in quark gluon plasma have been
studied in ref \cite{7}, where the results were reported
for a chemically equilibrated baryon-free plasma. It was shown that the
dissipative effect enhances the entropy production leading
to a dynamical path intermediate between idealized isentropic
and isoergic ones.
The energy density and temperature of the plasma also decrease at slower rate with
the inclusion of dissipative effects. This would lead to greater yields of direct
probes such as photons and dileptons, which are sensitive to the thermal history
of the reaction.
However, the hydrodynamic expansion for a viscous quark gluon plasma
undergoing chemical equilibration either at zero or at finite baryon density
has not been investigated so far.

  In this work, we study the dissipative effects in quark gluon plasma with
an emphasis on the results at finite baryon density. We derive a set of hydrodynamical
equations for the evolution of the energy-momentum tensor and parton densities
based on simple scaling laws. This set of equations has been applied to study
the hydrodynamical evolution of the viscous plasma undergoing chemical
equilibration. It is found that due to viscous heating, the temperature evolves
slower, whereas the parton fugacities evolve faster than the  ideal plasma
(without viscosity). This is contrary to the expectation that chemical
equilibration consumes more energy which makes the plasma cool faster.

The present work has been organized as follows. In  section II, we
begin with a brief review of the dissipative hydrodynamics and derive
a set of hydrodynamical equations using Landau definition of rest frame.
Using this set of equations,
the chemical equilibration of a viscous  baryon-free plasma is studied
in section III.
Finally, in section IV, we discuss the results and present the conclusions.
In the derivations, we have used  simple scaling law for the
dissipative corrections, for which a more formal justification has been
given in Appendix A using Boltzmann equation for the evolution of the parton
densities. The viscosity coefficient
is shown to be proportional to $\epsilon/ T$ which is also consistent with the
finite temperature QCD calculations as shown in Appendix B.

\section{HYDRODYNAMICS FOR A DISSIPATIVE SYSTEM}

Energy  dissipation  in a moving fluid occurs  due to internal friction (viscosity)
and  heat  exchange between different parts (thermal conduction) of the fluid.
If the mean free path of the constituent particles  is  comparable
to  the fluid dimension, the collective quantities like pressure,
energy density, number density, velocity etc.  may  vary  over  a
distance  of  mean  free  path.  This will lead to decay of fluid
kinetic energy as heat energy.
In nucleus-nucleus collisions, as the dimension of the plasma
is not very large compared to the mean free path of the partons,
dissipative effects may be quite important \cite{7}.

In the presence  of  dissipation,  the  energy-momentum  tensor $T^{ik}$ and
number density $n^i$   can be decomposed into an ideal and a dissipative
part \cite{8,9}
\begin{equation}
T^{i  k}=[(\epsilon+P)u^i  u^k  - Pg^{i k}]+\tau^{i k},
\end{equation}
\begin{equation}
n^{i }=nu^i + \nu^{i}.
\end{equation}
where  $\epsilon$, P and n  are  the  energy  density,  pressure   and
particle number density and $u^i=\gamma(1,\vec{v})$ is four-velocity in
terms of local fluid velocity $\vec{v}(x)$ and $\tau^{ik}$ and $\nu^i$
are the dissipative corrections
\paragraph*{}
The relativistic hydrodynamics is  based on the local conservation laws:
\begin{equation}
\partial_i T^{ik}=\partial_i n^i=0\label{emcons}.
\end{equation}
In the present context $n^i$ represents the baryon four vector and the
equation $\partial_i n^i$=0 signifies the conservation of baryon number.
However, if there is no net particle production i.e. chemical equilibration
is complete, the above equation is  applicable to the individual partons as well.

The form of the dissipative terms $\tau^{ik}$ and $\nu^i$ depend on the
definition of what constitutes the local rest frame of the fluid. The
four velocity $u^i$ should be defined such a way that in a proper frame
of any given fluid element, the energy and the number density are expressible
in terms of other thermodynamic quantities by the same formulae as when
dissipative processes are not present. It is also necessary to specify whether
$u^i$ is the velocity of energy transport or particle transport. Accordingly,
there exists two definitions for the rest frame; one due to Landau and other
due to Eckart. In Landau approach, $u^i$ is taken as the  velocity of the
energy transport so that energy three flux $T^{0i}$ vanishes in a comoving frame.
The Landau definition requires $u^i \tau^{ik}=u_i\nu^i=0$ so that
$T^{00}=\epsilon$ is the energy density and $n^0=n$ is the number density
in the proper frame of the fluid. In the Eckart definition, $u^i$ is
taken as the velocity of the particle transport  and the particle three current,
rather than the energy three flux vanishes in the fluid rest frame. So in the
Eckart definition of rest frame, the particle four vector can be written as
$n^\mu=(n,{\vec 0})$,
whereas in the Landau definition of rest frame $n^\mu=(n,{\vec \nu})$. Therefore,
the two frames are related by a Lorentz transformation with a boost velocity ${\vec \nu}/n$.

It is found  that due to ill defined boost velocity \cite{7},
the energy three flux in the Eckart frame, which involves heat conductivity $\kappa$ is not well
defined as it diverges in the limit of chemical potential $\mu \rightarrow$0.
On the otherhand, in the
Landau definition heat conduction enters as a correction to baryon flux.
It was shown that inspite of the divergence of $\kappa$, the correction to
the baryon flux $\nu^i$ is finite \cite{7} . We  use  the Landau
definition for the subsequent study of the evolution of the energy momentum
tensor and parton densities of the quark gluon plasma.

Under the assumption that the dissipative terms $\tau^{ik}$ and $\nu^i$
are of first order in the gradients, the requirement that the entropy
increases with time ( $\partial_\alpha \sigma^\alpha > 0$,
~~~$\sigma^\alpha = \sigma u^\alpha - \mu \nu^\alpha /T$ with  chemical
potential $\mu$ and  temperature T )
leads to (in the Landau definition) \cite{7,8}
\begin{equation}
\tau^{\alpha    \beta}=\eta(\nabla^\alpha    u^\beta+\nabla^\beta
u^\alpha-\frac{2}{3}\triangle^{\alpha \beta} \nabla_\rho u^\rho)
+\xi \triangle^{\alpha \beta} \nabla_\rho u^\rho,
\end{equation}
\begin{equation}
\nu^\alpha=\kappa\left[\frac{nT}{\epsilon+P}\right]^2
\nabla^\alpha\left[\frac{\mu}{T}\right].
\end{equation}
where $\triangle^\alpha _\beta=g^\alpha_\beta-u^\alpha u_\beta$ and $\nabla^
\alpha=\triangle^{\alpha \beta}\partial_\beta$, $\eta$ and $\xi$ are the
co-efficients of the shear and bulk viscosities and $\kappa$ is the
co-efficient of heat conduction. As mentioned before, the heat conductivity
does not enter in the energy flux $T^{0i}$, but rather as a finite baryon current
$\nu^i \propto \partial^i(\mu/T)$ in the rest frame of the fluid.
In the presence
of dissipation, the conservation equations (3) need to be solved with the
above correction terms. Using one-dimensional boost-invariant scaling law \cite{1}
and Eq. (4) for $\tau^{ik}$ the identity $u_k \partial_i T^{ik}$=0,
resulting from the energy momentum conservation,  takes the simple form \cite{7}
\begin{equation}
\frac{\partial \epsilon}{\partial \tau}+\frac{\epsilon+P}{\tau}=
\frac{4}{3}\frac{\eta+\xi}{\tau^2}.
\end{equation}
The evaluation of $\partial_i n^i=0$ is not straightforward as one needs to know the
space time dependence of $\mu/T$ and also care has to be taken so that $\nu^i$
is finite although $\kappa$ diverges in the limit $\mu \rightarrow 0$ \cite{7}.
However, we will follow a simple procedure to obtain the evolution of the
baryon density directly without invoking the  baryon current
(hence the conductivity) as given by Eq. (5).
It may be mentioned here that
in addition to the
baryon number conservation, the parton densities (quark, anti-quark and gluon)
also evolve satisfying the relation $\partial_i n^i=0$ in absence of any particle
production. Clearly,
the evolution of the parton densities will also get modified due to the
dissipative corrections.
We estimate these dissipative corrections in a simple
way from the following considerations. Since the shear viscosity
scales as $T^3$, while the bulk viscosity $\xi$ vanishes for a quark gluon
plasma, one can define an acceptable range for the shear viscosity
$\eta$ given by \cite{7}:
\begin{equation}
2T^3\le \eta \le \frac{1}{4}\epsilon \tau. \nonumber
\end{equation}
so that Navier-Stokes scaling theory is applicable to the expansion of the
plasma. A baryon-free non-viscous plasma ($\mu$=0, $\eta$=0)
leads to $\epsilon \tau^{4/3}$=$\epsilon_0 \tau_0^{4/3}$.
Using the upper bound of viscosity as $\eta$=$\epsilon \tau/4$ and substituting
$\epsilon = 3P $ in Eq.(6), we get $\epsilon \tau$=$\epsilon_0 \tau_0$.
This means that T scales as $\tau^{-1/3}$ for $\eta=0$ and as $\tau ^{-1/4}$
for $\eta$ approaching the upper bound.  Therefore, it is reasonable to
assume that T  scales as $\tau^{-\gamma}$  for any  value of $\eta$ in the range
$1/3 \le \gamma \le 1/4$. Since $\epsilon$ scales as $\tau^{-4\gamma}$,
the LHS of Eq.(6) can be expressed  in terms of $\gamma$ as,
\equation
\frac{\partial \epsilon}{\partial \tau}+\frac{\epsilon+P}{\tau}=
 \frac{4}{3}(1-3\gamma)\frac{\epsilon}{\tau}.
\endequation
On similar grounds,  the evolution of parton density $n_i$ ( i=q, $ {\bar q}$, g )
can be described in one dimension as (since $n_i$ scales as $\tau^{-3\gamma}$),

\begin{equation}
\frac{\partial n_i}{\partial \tau}+\frac{n_i}{\tau}=
 (1-3\gamma)\frac{n_i}{\tau}.
\end{equation}
For an ideal plasma, $\gamma =1/3$ and the RHS of Eqs. (8) and (9) vanish,
while for  other values of $\gamma$ the RHS is non-zero which can be interpreted
as the correction due to dissipation.
Comparing RHS of Eq.(8) with Eq.(6), we can identify
$$\eta=(1-3\gamma)\epsilon \tau.$$
It is also interesting to note that $\eta=\epsilon \tau/4$
for $\gamma =1/4$ is consistent with the upper limit of viscosity. In terms of $\eta$,
Eq. (9) can be written as
\begin{equation}
\frac{\partial n_i}{\partial \tau}+\frac{n_i}{\tau}=
\frac{\eta n_i}{\epsilon \tau^2}.
\end{equation}
The dissipative correction appearing in Eq. (10) is derived for a baryon-free plasma.
However, as shown in Appendix A, the above correction is also valid
for $\mu \ne 0$ if $\epsilon$ and $n_q({\bar q})$ are
defined appropriately for a baryon-rich plasma. Finally, the evolution of baryon
density can be obtained from Eqs.(9) and (10) as
\begin{equation}
\frac{\partial n_B}{\partial \tau}+\frac{n_B}{\tau}=
 (1-3\gamma)\frac{n_B}{\tau}~~=\frac{\eta n_B}{\epsilon \tau^2}.
\end{equation}
where $n_B=n_q-n_{\bar q}$. It may be noted here
that although the baryon conductivity vanishes for $\mu=0$, the
dissipative corrections for the quark and anti-quark densities are finite as
seen from Eq.(10). It may be mentioned here that for a baryon-symmetric matter,
 the energy momentum equation given by Eq.(6) is sufficient to describe the
dynamical evolution of the plasma. However, Eq.(10) has been derived for consistency
and will be useful to study the density evolution in presence of particle
production particularly when the chemical equilibration is not complete.

The dynamical evolution of a baryon-rich plasma can be studied by solving
Eqs.(6) and (11) numerically. For the plasma phase, we consider only the u and d
quarks for which the energy and baryon densities are given by
\begin{equation}
\epsilon=T^4(\frac{37\pi^2}{30} + 3 x^2 +\frac{3}{2\pi^2}x^4),
\end{equation}
\begin{equation}
n_B=\frac{2}{3}T^3(x+\frac{x^3}{\pi^2}).
\end{equation}
where $x=\mu/T$. The shear viscosity $\eta$ depends on both T and $\mu$
in a complicated way \cite{11}. However, as shown in Appendix B , we
can approximate $\eta \approx \eta_0
\epsilon/T$ which scales as $T^3$ in case of a baryon free plasma.
It is found that $\eta_0$ weakly depends on T and $\mu$. Therefore, we
will treat it as a constant parameter for the subsequent calculations.

Figures. 1(a) and 1(b) show the time evolution of the ratio of energy density
$\epsilon/\epsilon_0$ and total entropy $( \sigma \tau )/ ( \sigma_0 \tau_0 )$
for a plasma at
 $x_0(\mu /T)$=1.0, energy density $\epsilon_0$= 9.0 GeV/$fm^3$ (corresponding
to an initial temperature of $T_0$= 0.258 GeV ) with different values of viscosity
coefficients $\eta_0$= 0.0, 0.5 and 1.0 . As seen from the Figures 1(a) and (b), with
increase in viscosity, the energy
density evolves faster than that corresponding to the isentropic expansion
($\eta_0=0$) and also extra entropy is produced when the medium is viscous.
Figure 1(b) also shows that most of the entropy is produced in the
first few fm/c and the asymptotic value of $\sigma \tau$ is approached quickly.
So these results are  consistent with the
findings of \cite{7}  except that the present work has been extended to finite
baryon density. As mentioned before, the shear viscosity
may have an upper bound as given by Eq. (7). However, in the above calculations, we only
carry out a parametric study to see the effect of viscosity on the evolution
of the energy density
and entropy production although some of the $\eta_0$ values may exceed the upper
limit.

Figures 2(a) and 2(b) show the variation of $T/T_0$ and $\sigma \tau/(\sigma_0
\tau_0)$ as a function of $\tau$
for different baryon densities of $x_0$=0.0, 1.0 and 2.0 at fixed
$\eta_0$=1.0 and $\epsilon_0$=9.0 GeV/fm$^3$. For a given temperature T,
the viscosity of the plasma at finite baryon density is higher than the
baryon-free case (since $\epsilon$ increases with $\mu$). This is  true
even if the energy density is held fixed since the plasma
will have lower initial temperature at finite baryon density
(resulting in  higher viscosity).
Since the viscosity increases with $x$, the
temperature T evolves slower and more entropy is produced as compared to the
baryon free case. It is also interesting to note that whatever be the value
of viscosity, both T and $\mu$ scale with the same powerlaw as
$\tau^{-\gamma}$
(see the solid circles in Figure 2a for $\mu/\mu_0$ which overlap with the
curve for $T/T_0$.) This means
the ratio of $\mu$  and T (i.e. $x$) remains independent of $\tau$,
consistent with the assumption made in appendix A.
This is true even in case of an ideal ($\eta_0=0$) baryon-rich plasma where both $\mu$
and T scale as $\tau^-{1/3}$ i.e. both obey Bjorken's scaling.
However, we have found
that $x$ decreases with $\tau$ if the hydrodynamical expansion proceeds along
with the chemical equilibration \cite{12}. The $\tau$ dependence of the chemical
potential $\mu$ and the temperature T will be different due to parton production
during chemical equilibration.

\section { EFFECT OF VISCOSITY ON CHEMICAL EQUILIBRATION FOR BARYON FREE PLASMA}
The plasma in equilibration means there is no net parton production and
the momentum distribution of the partons can be described either by Fermi-Dirac (FD) or
by  Bose-Einstein (BE) distribution function.
The picture emerging from the numerical simulations of parton cascades supports
the formation of a plasma at RHIC and LHC energies which is initially gluon-rich.
Rapid gluon scattering at a high initial temperature will lead to
substantial thermalization of the plasma in a very short time while it will
be far off from chemical equilibrium.
Therefore, process of chemical equilibration
will proceed along with the dynamical evolution of the plasma. The process of
chemical equilibration will accelerate the
cooling of the plasma further and the plasma may not achieve equilibrium by the time
temperature drops to the critical value \cite{4,5,6}. However, all theses calculations
assume an ideal plasma where the effects of dissipative
processes are neglected.
The chemical evolution of a viscous plasma will be
different than its ideal counterpart due to extra entropy production
and viscous heating of the plasma.

The distribution functions for quarks (anti-quarks) and gluons
appropriate for a plasma which has achieved kinetic equilibrium but
is far off from chemical equilibrium are given by
\begin{equation}
f_ q(_{\bar q})  = \frac{\lambda_ q(_{\bar q})} {1 + e^{(p\mp \mu )/T}}
~~~~;~~~~ f_ g = \frac{\lambda_ g}{e^{p/T}-1}.
\end{equation}
where $\lambda_i $'s  give the measure of
the deviation of the distribution functions from the equilibrium values.
The chemical equilibrium is said to be achieved when $\lambda_i \rightarrow 1$
and the baryo-chemical potential $\mu$ attains an equilibrium value $\mu_{eq}$
consistent with the baryon current conservation.
It may be mentioned here that  although the plasma is dissipative, we
use the FD and BE distribution functions for quarks, anti-quarks and gluons.
However, the temperature and the chemical potential will evolve as per
viscous hydrodynamics and will differ from their ideal counter parts. As
discussed in Appendix A, the present formalism of viscous hydrodynamics is not
valid for a baryon rich plasma undergoing chemical equilibration. Since $x$ depends
on $\tau$, additional corrections are needed in the rate equations
describing the evolution of the viscous plasma. Therefore, in the following,
we have studied the process of chemical equilibration
for a viscous plasma only at zero baryon density.

We restrict our considerations
to the following two reaction mechanisms for the equilibration of the parton
flavours $gg \rightleftharpoons ggg$ and $gg \rightleftharpoons q \bar q$.~~
The evolution of the parton densities are governed by the master equations
\begin{equation}
\partial_\mu (n_g^\mu) = (R_{2\rightarrow 3} -R_{3\rightarrow 2})
-(R_{g\rightarrow q} - R_{q\rightarrow g}),
\end{equation}
\begin{equation}
\partial_\mu (n_q^\mu) = \partial_\mu (n_{\bar q}^\mu)=
(R_{g\rightarrow q} - R_{q\rightarrow g}).
\end{equation}
where $R_{2\rightarrow 3}$ and $R_{3\rightarrow 2}$ denote the rates for the
process
$gg \rightarrow ggg$ and its reverse, and $R_{g \rightarrow q}$ and $R_{q \rightarrow g}$
are for the process $gg \rightarrow q \bar q$ and its reverse respectively.
Note that the density four vector $n^i=n u^i+\nu^i$ now includes a
dissipative correction $\nu^i$.
The rate equations are in general quite complicated to solve.
For baryon-free plasma ($\mu$=0),
the authors in \cite{4}, have used a simple factorization for the RHS of the above equations
based on a classical approximation i.e. using the Boltzmann distribution
function for quark, anti-quark and gluon and eliminating the Pauli blocking
and Bose enhancement factors in the final states. However, it was shown  by us
recently \cite{12} that the same factorization can be used, even by including
full quantum statistics, particularly when the plasma is  highly unsaturated.
Using the viscous hydrodynamics as derived in the previous section and with the
factorization given in Ref \cite{4}, the Eqs. (15) and (16) can be written as
\begin{equation}
\frac{\partial n_g}{\partial \tau} +\frac{n_g}{\tau} =
n_gR_3(1-\lambda_g)-2n_gR_2(1-\frac{\lambda_q \lambda_{\bar q}}{\lambda_g^2})
+\frac{n_g\eta}{\epsilon \tau^2},
\end{equation}
\begin{equation}
\frac{\partial n_q}{\partial \tau}+\frac{n_q}{\tau}=
n_gR_2(1-\frac{\lambda_q \lambda_{\bar q}}{\lambda_g^2})
+\frac{n_q\eta}{\epsilon \tau^2},
\end{equation}
\begin{equation}
\frac{\partial n_{\bar q}}{\partial \tau}+\frac{n_{\bar q}}{\tau}=
n_gR_2(1-\frac{\lambda_q \lambda_{\bar q}}{\lambda_g^2})
+\frac{n_{\bar q}\eta}{\epsilon \tau^2}.
\end{equation}

The above set of equations differ from the corresponding equations of ref.[4]
with the last terms containing the viscosity factors.
The density weighted cross sections $R_3$ and $R_2$ are given by \cite{4}
\begin{equation}
R_3 \approx 2.1 ~\alpha_s^2 ~T ~\sqrt{2\lambda_g-\lambda_g^2}~~~~~;~~~
R_2 \approx .24 N_f~\alpha_s^2~ \lambda_g ~T ~ln(\frac{1.65}{\alpha_s \lambda_g}).
\end{equation}
Using the distribution functions given by (14)  for quarks, anti-quarks and the
gluons, the energy and number densities can be written as:

\begin{equation}   \epsilon=T^4[a_2\lambda_g+b_2(\lambda_q +\lambda_{\bar q} )].
\end{equation}

   with $a_2=8 \pi^2/15$;~~ $b_2=N_f(7 \pi^2/40)$~; where $N_f$ is the
   dynamical quark flavours. Similarly, the number densities
   for gluon, quark and antiquark are:

\begin{equation}  n_g=\lambda_g a_1 T^3 ;~ a_1=\frac{16}{\pi^2}~~ \zeta (3), \end{equation}
\begin{equation}  n_q=\lambda_q b_1 T^3 ;~ b_1=\frac{9}{2\pi^2}~~ \zeta (3) N_f, \end{equation}
\begin{equation}   n_{\bar q}=\lambda_{\bar q} b_1 T^3 . \end{equation}

The above set of Eqs. (17-19) along with Eq. (6) were solved
using fourth order Runga-Kutta method, to obtain the time dependence of the
temperature T, the quark and anti-quark fugacities $\lambda_g$,
$\lambda_q$ (= $\lambda_{\bar q}$).
The initial conditions were taken from the HIJING calculations both
for RHIC and LHC energies \cite{4}. We take
$T_0$= 0.57 GeV, $\lambda_{g0}$=0.09, $\lambda_{q0}$ =0.02
at $\tau_0$ =0.31 fm for RHIC energy
and T=0.83 GeV, $\lambda_{g0}=$0.14, $\lambda_{q0}$=0.03 and $\tau_0$=0.23 fm
for LHC energy. Since, energy density depends on fugacities and chemical
potential, viscosity will also depend on them implicitly.

Figures (3) and (4) show the variation of T, $\lambda_g$ and $\lambda_q$
with proper time $\tau$ for $\eta_0$=0.0 and 0.2. As mentioned above,
these calculations correspond to a non-equilibrated plasma at
zero baryon density. In the absence of dissipation, i.e. $\eta_0$=0, the plasma cools
faster than that predicted by the Bjorken scaling (see the curves with solid
circles in Figure 3(a) and 4(a)~), since additional energy is consumed in
in the process of chemical
equilibration. A further rise in equilibration rate (i.e rise in fugacities) will
result in faster cooling of the plasma. The reverse happens in presence
of viscosity. As seen from the Figures (3) and (4), during chemical equilibration,
the quark and gluon equilibration rates become faster
when the medium becomes more viscous. This, however, does not result
in a faster cooling of the plasma as expected.
So in the case of a viscous plasma,
the chemical equilibration becomes faster and the temperature
drops more slowly due to generation of  heat.
Since the presence of viscosity makes the equilibration faster and also
increases the life time of the plasma phase, it would be interesting to see
whether chemical equilibration can be achieved when the medium is viscous.
While this requires the precise knowledge of viscosity , the short-dashed curve in
Figures (3) and (4) shows the result that can achieved with the upper limit
of viscosity $\eta=(\epsilon\tau)/4$ ( i.e $\eta_0=\frac{T\tau}{4}$). These
calculations show that even for the limiting case, the temperature may still drop
faster than the Bjorken scaling and the plasma may not achieve chemical
equilibration  by the time T drops to $T_c \approx 0.2$ GeV, provided
the initial conditions taken from HIJING calculations are prevalent at RHIC
and LHC energies.

 The above results are obtained for the chemical evolution of a viscous plasma
 at zero baryon density. In a recent paper, we have shown that \cite{12} in case of
a non-viscous plasma at finite baryon density, the rate of chemical
equilibration slows down particularly for gluons resulting in slower cooling
of the plasma. Therefore, in the presence of baryon density, the cooling rate
of the viscous plasma is expected to  slow down further. However, it is
difficult to conclude about the net gluon equilibration rate
since it is suppressed at finite baryon density.

\section {SUMMARY AND CONCLUSION}
We have studied the properties of a viscous quark gluon plasma at finite
baryon density. A set of equations that describes the space time evolution
of the energy-momentum and parton densities has been derived based on a
simple scaling law as well as by solving Boltzmann approximation under boost
invariant scenario. The total entropy is conserved in case of an ideal plasma
and the temperature and chemical potential evolve as per the Bjorken's scaling.
However, for a viscous plasma additional entropy is produced due to viscosity
both at zero and finite baryon density. In fact, finite baryon density makes
the plasma more viscous as energy density increases with $\mu$. Even at a fixed energy
density, viscosity rises due to the lower initial temperature. Therefore,
plasma at finite baryon density  produces more entropy as compared to the baryon
free case. We have also studied the chemical equilibration of the plasma at zero
baryon density using the rate equations appropriate for a viscous plasma.
It is found that due to viscosity , the temperature evolves more slowly,
while the fugacities evolve faster than the ideal plasma.
The slow cooling of the plasma and the fast equilibration rates of
gluon and quark (anti-quark) will affect the thermal photon and di-lepton
yields which are the important probes to study the  dynamical evolution
of the quark gluon plasma.
The thermal open charm production will also be enhanced particularly at high
$P_T$ or at high invariant mass regions. The additional entropy which is
generated due to viscosity will also reflect on an increased multiplicity
distributions of the final particles. Further studies in this direction will be
important to understand the behavior of the dynamical evolution of the quark
gluon plasma produced in relativistic heavy ion collisions.

To conclude, larger transverse momentum associated
with the hydrodynamic expansion would be reduced as collective flow velocities
are dissipated into heat. The dissipative effects may dampen the
fluctuations which otherwise are expected to serve as signatures of unusual phenomena of QGP
phase transitions. Since most proposed observables of the plasma are sensitive to the full
space-time history of the reaction, dissipative phenomena must be taken into
account if quantitative predictions are to be made. In future, we plan to
estimate the dilepton and photon yields from QGP phase with the inclusion of these
dissipative effects.
\section*{Acknowledgement} We are greatful to Prof. Larry McLerran
for his useful comments and guidance.
\setcounter{equation}{0}
\section *{Appendix A}
\renewcommand{\theequation}{A.\arabic{equation}}

The dissipative corrections $\tau^{ik}$ and $\nu^i$ in Eqs. (4) and (5) can be
evaluated using the Boltzmann equation $p^\mu\partial_\mu f_a=C_a(f)$, which
describes the evolution of the Wigner densities $f_a(x,p)$ of the parton of
type $a$ in terms of the collision integrals $C_a$.
At equilibrium, the collision integral C(f) vanishes and the distribution
function $f_0(p,\mu,T)$ becomes Bose-Einestein or
Fermi-Dirac type. However, in the presence of dissipation, the hydrodynamical
solution assumes only local equilibrium where
\begin{equation}
f_H=f_0(p_\mu u^\mu (x), \mu(x), T(x)).
\end{equation}
with the four velocity u, chemical potential $\mu$ and temperature T
being functions of $x_\mu$. However, $p^\mu \partial_\mu f_H \ne 0$, while
$C(f_H)=0$. So $f_H$ cannot be a solution of the Boltzmann equation and there has to be a
correction to $f_H$ by an amount $\delta f$ that is first order in gradients
of u, $\mu$ and T. Writing  $f=f_H\pm\delta f$, the Boltzmann equation can
be written as
\begin{equation}
p^\mu \partial_\mu f_H =  p^\alpha u_\alpha \frac{\delta f}{\tau_c}.
\end{equation}
where $\tau_c$ is the relaxation time which  depends on time . In terms of
$\delta f$, the dissipative corrections are given by \cite{7}
\begin{equation}
\tau^{ik} = \int \frac{d^3p}{{(2\pi)}^3} \frac{p^i p^k}{p^0}
(\delta f_g + \delta f_q +\delta f_{\bar q}),
\end{equation}
\begin{equation}
\nu^i = \int \frac{d^3p}{{(2\pi)}^3} \frac{p^i}{p^0}
(\delta f_q -\delta f_{\bar q}).
\end{equation}
The Boltzmann equation (A.2) in (1+1) dimension has the simple form:
\begin{equation}
\frac{\partial f_H}{\partial t}+v_{z}\frac{\partial f_H}{\partial z}
=\frac{\delta f}{\tau_c}\label{boltz}.
\end{equation}
where $v_z$ is the parton velocity.
As in \cite{13}, we further assume that the central rapidity regime is initially approximately
Lorentz invariant under longitudinal boost so that the conditions at point z at
time t are the same as those at $z=0$ at proper time $\tau=(t^2-z^2)^{1/2}$.
Hence the left-hand side of Eq. (\ref{boltz}) at $z=0$ becomes
simply the time derivative at constant $p_zt$ \cite{13}
\begin{equation}
\left.\frac{\partial f_H(p_\bot,p_z,t)}{\partial t}\right|_{p_z t}
=\frac{\delta f(p_\bot,p_z,t)}{\tau_c}.
\end{equation}
Therefore, the integral equations A(3) and A(4) can be evaluated using
$\delta f$ given by
\begin{eqnarray}
\delta f&=&\tau_c
\left.\frac{\partial f_H(p_\bot,p_z,t)}{\partial t}\right|_{p_z t}\nonumber\\
&=-&\tau_c\left.f_0\bar{f_0}\frac{\partial }{\partial t}\left(\frac{p}{T}\right)\right|_{p_zt}.
\end{eqnarray}
where $\bar f_0=(1\mp f_0)$ for fermions and bosons.
An ideal plasma in one dimension expands
following the
Bjorken's scaling where the temperature T decreases as $t^{-1/3}$.
In presence of dissipation, there is an extra entropy production and the plasma
cools slowly. However, we can still assume a power law dependence for $T\propto
t^{-\gamma}$ where $1/4\le\gamma \le 1/3$. Thus, the time derivative of (p/T)
can be evaluated keeping $p_z t$ constant
\begin{equation}
\delta f=\tau^c~f_0~\bar f_0 \frac{p}{T t}(\cos^2 \theta-\gamma).
\end{equation}
where $p_z=p \cos \theta$ and $p^2=p_\bot^2+p_z^2$ . Here
we have assumed a baryon-free plasma ( $\mu=0$).
However, for finite baryon density, p in Eq. (A.7)
needs to be replaced by $(p\pm\mu)$ for anti-quark
or quark distribution function respectively.
Consequently, there will be an extra contribution arising from the term
$\partial x/\partial t$
where $x=\mu/T$. In case of a non-viscous plasma,
by solving the equations
for the conservation of entropy ($\sigma$t=const) and  of
baryon number ($n_B$t=const), it can be shown that both $\mu$ and T decrease
as $t^{-1/3}$ even though the plasma is  baryon-rich. This implies that  the ratio $x$ remains independent
of t.  Therefore, it is reasonable to assume that $x$ will also remain
independent
of t even for a viscous plasma at finite baryon density.
This would mean  both $\mu$ and T should follow the same power
law as $t^{-\gamma}$.
In fact, this assumption is consistent with the
results that we find in section II (see solid circles for $\mu/\mu_0$ in figure 2a).
Since $x$ does not depend on t, there would not be
any extra correction and Eq.(A.8) can be used for $\delta f$ with
distribution function $f_0$ (for the quark and anti-quark)
defined appropriately for a plasma at finite baryon density. The above
assumption is not valid for a plasma undergoing chemical equilibration.
It is shown in a recent work \cite{12} that
$x$ decreases with t due to parton production during chemical equilibration,
although the time dependence can be neglected for small values of $x$.
Since $x$ varies with t, Eq. (A.8) is not valid in case of a
chemically equilibrating plasma
at finite baryon density except when $x$=0 or very small.

Using $\delta f$ given by Eq. A.8  and BE or FD distribution
for $f_0$, the dissipative corrections for baryon-rich plasma,
can be evaluated from Eqs. (A.3) and (A.4).
Consider the case $i = k =0$, for which $\tau^{00}$ and $\nu^0$ can be written as
\begin{equation}
\tau^{00} = \frac{1}{T t}\int \frac{d\phi}{{(2\pi)}^3}
\int p^4~(f_g{\bar f_g}\tau_c^g+f_q{\bar f_q}\tau_c^q+f_{\bar q}{\bar f_{\bar q}}\tau_c^{\bar q})
~dp~\int d\theta~ (cos^2\theta-\gamma)sin\theta ,
\end{equation}
\begin{equation}
\nu^{0} = \frac{1}{Tt}\int \frac{d\phi}{{(2\pi)}^3}
\int p^3~(f_q{\bar f_q}\tau_c^q -
f_{\bar q}{\bar f_{\bar q}}\tau_c^{\bar q})~dp~\int d\theta~
(cos^2\theta-\gamma)sin\theta .
\end{equation}
Replacing the $\theta$ integration by $(2/3)(1-3\gamma)$ and using
\begin{eqnarray}
\int_0^\infty p^n f_0 {\bar f_0} dp = n T \int_0^\infty p^{n-1} f_0 dp,
\nonumber \end{eqnarray}
Eqs. (A.9) and (A.10) can be written as

\begin{equation}
\tau^{00}=\frac{4\tau_c}{3t}(1-3\gamma)\epsilon,
\end{equation}
\begin{equation}
\nu^{0}=\frac{\tau_c}{t}(1-3\gamma)(n_q-n_{\bar q}).
\end{equation}

where $\epsilon = \epsilon_q+\epsilon_{\bar q}+\epsilon_g$ is the total
energy density and $n_q$ , $n_{\bar q}$ are the quark and anti-quark densities.
Similarly, it can be shown that $\tau^{0z}$ and $\nu^z$ will vanish due to
vanishing $\theta$ integration. Recall that the terms which need to be added
to the hydrodynamic Eqs.(3) are $u_k \partial_i \tau^{ik}$ and
$\partial_i \nu^i$. Since, we consider the expansion in (1+1) dimension
around the central rapidity regime at z=0, the terms which are of interest
to us are $\partial_0 \tau^{00}$ and $\partial_0 \nu^0$.
Although, gluons equilibrate much faster than the quarks and anti-quarks, we
will assume an effective relaxation time for gluon, quark and anti-quark given
by $\tau_c^{-1} \propto T \alpha_s^2 ln(\alpha_s^{-1})$ \cite{14}. Neglecting
time dependence of $\alpha_s$ one can have
\begin{equation}
\partial_0 \tau^{00}= -\frac{4}{3}(3 \gamma +1)(1- 3\gamma)\frac{\epsilon}{Tt^2} ,
\end{equation}
\begin{equation}
\partial_0 \nu^0= -(2 \gamma +1)(1- 3\gamma)\frac{n_B}{T t^2} .
\end{equation}
Therefore, in the presence of dissipation ,  terms proportional to
$\epsilon/(Tt^2)$ and $n_B/(Tt^2)$ are to be added
to the Eq. (3) describing the evolution of the energy momentum
and parton densities. It is also interesting to note that for an ideal
 plasma $\gamma=1/3$ so that the dissipative correction vanishes.
However, comparing Eq. A(13) with RHS of Eq. (6) which has been derived
from Eq.(4) directly, we can write
\begin{equation}
\partial_0\tau^{00} =- \frac{4}{3} \frac{\eta}{t^2} . \nonumber
\end{equation}
where $\eta=(3\gamma+1)(1-3\gamma)\epsilon/T$.

Similarly, in terms of $\eta$, we can have
\begin{equation}
\partial_0\nu_0 =-C \frac{n_B\eta}{\epsilon t^2} . \nonumber
\end{equation}
Therefore, we can write the following evolution equations in terms of $\eta$
\begin{equation}
\frac{\partial \epsilon}{\partial t}+\frac{\epsilon+P}{t} =
\frac{4}{3}\frac{\eta}{t^2} ,
\end{equation}
\begin{equation}
\frac{\partial n_B}{\partial t} +\frac{n_B}{t} =
\frac{C n_B\eta}{\epsilon t^2}.
\end{equation}
where $C=(2\gamma+1)/(3\gamma+1)$. It is also possible to write the density
evolution for quarks and anti-quarks separately as:
\begin{equation}
\frac{\partial n_q}{\partial t} +\frac{n_q}{t} =
\frac{C n_q\eta}{\epsilon t^2} ,
\end{equation}
\begin{equation}
\frac{\partial n_{\bar q}}{\partial t}+\frac{n_{\bar q}}{t} =
\frac{C n_{\bar q}\eta}{\epsilon t^2} .
\end{equation}
The above equations have the same form as that of Eqs. (6), (10) and (11).
The value of C lies between 0.83 and 0.86 depending on the value of $\gamma$ .
Ideally, we expect C=1 from the requirement that
in the absence of baryon density all the Eqs. A(17)-A(20) should give
same solution for temperature.
This deviation ($C < 1$) arises due to various approximations that are used
in deriving $\delta f$. For example, we have used an effective relaxation
time $\tau_c$ both for quarks and gluons although $\tau_q > \tau_g$. Similarly,
we use $T = A t^{-\gamma}$ where A is taken as constant, although it may depend
on t. However, the deviation is not too big and we will use C=1 for
consistency in the subsequent studies.
It is also important to note that the form of viscosity
$\eta \propto \epsilon/T$ is also consistent with the results of \cite{11}
which was obtained using finite theory QCD (see Appendix B).

\section *{Appendix B}

\setcounter{equation}{0}
\renewcommand{\theequation}{B.\arabic{equation}}
The shear viscosity co-efficient using the relativistic kinetic
theory for a massless QGP under relaxation time approximation
can be written as \cite{11,15}
\begin{equation}
\eta_i=\frac{4}{15}\epsilon_i\lambda_i .
\end{equation}
where $\epsilon_i(i=q,\bar q,g)$ is the energy density of particle type $i$
and $\lambda_i$ is the mean free path, which is the inverse interaction
rate, $\lambda_i=1/\Gamma_i$.
Recently, Hou and Jiarong \cite{11}  have evaluated the interaction rates for
quarks and gluons as
for baryon rich plasma using finite temperature QCD
\equation \Gamma_q=T\nu,~~~~~~~~\Gamma_g=\frac{9}{4} T \nu. \endequation
with
\begin{equation}
\nu=\frac{16\pi}{27}\alpha_s^2\left(\frac{4}{3}+\frac{1}{\pi^2}
x^2\right)\left(-\ln\alpha_s+\ln\left(\frac{1}{4\pi(\frac{4}{3}+\frac{1}{\pi^2}
x^2)}+\alpha_s\right)+D\right) .
\end{equation}
where D is the gluon damping factor given by
\begin{equation}
D\approx\frac{N_c^2\alpha_s}{4\pi\left(\frac{4}{3}+\frac{1}{\pi^2}x^2\right)}
\ln\left(\alpha_s^2\frac{N_c^2}{1+N_c^2\alpha_s^2}\right) .
\end{equation}
Finally, the viscosity co-efficient $\eta$ is given by
\begin{equation}
\eta_q=\frac{\epsilon_q}{T\nu},~~~~~~~~~~~\eta_g=\frac{4\epsilon_g}{9T\nu} .
\end{equation}
Therefore, the $\mu$ dependence of the viscosity arises through $\epsilon_q$
and $\nu$.
We have studied $\nu$ as a function of $\mu$ at various temperatures T
using the following expression for the running coupling constant:
\begin{equation}
\alpha(T,\mu)=\frac{12\pi}{(33-2N_f)\ln\left(T^2\frac{0.8x^2+15.622}
{\Lambda_s^2}\right)} .
\end{equation}
where $\Lambda_s$=0.1 GeV and $N_f$=2.
It is found that $\nu$ does not depend on $\mu$ very strongly. Even the variation
of $\nu$ with temperature is also not very significant. The viscosity primarily
depends on $\epsilon/T$ where $\epsilon$ is the total energy density of the
plasma. Therefore, we can write
\begin{equation}
\eta = \eta_0 \frac{\epsilon}{T} .
\end{equation}
where $\eta_0$ is treated as a constant although it may depend on T and $\mu$
rather weakly.

\begin{figure}[t!]
\vspace{3cm}
\hspace{3cm}
\begin{minipage}[t]{12cm}
\psfig{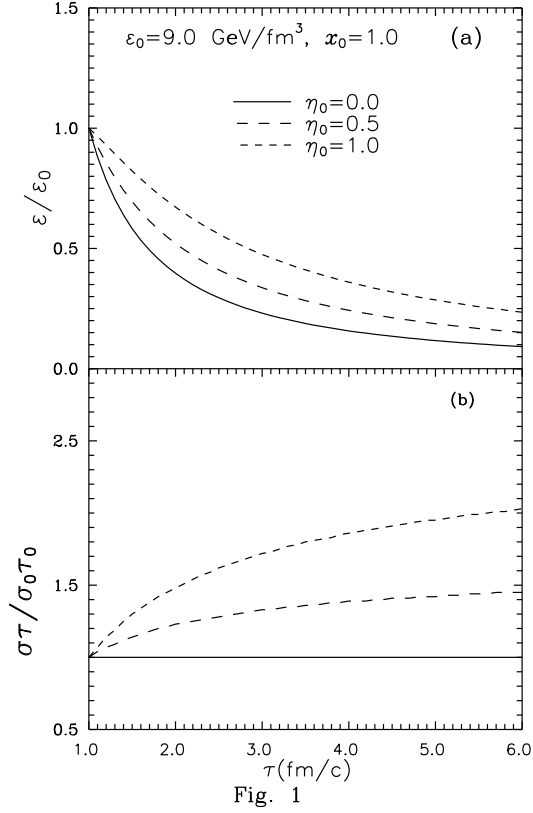}
\caption{ (a) The ratio of the energy density $\epsilon / \epsilon_0$ ,
and (b)the ratio of the total entropy $(\sigma \tau)/(\sigma_0\tau_0)$
as a function of proper time $\tau$.}
\label{Figure 1}
\end{minipage}
\end{figure}
\newpage
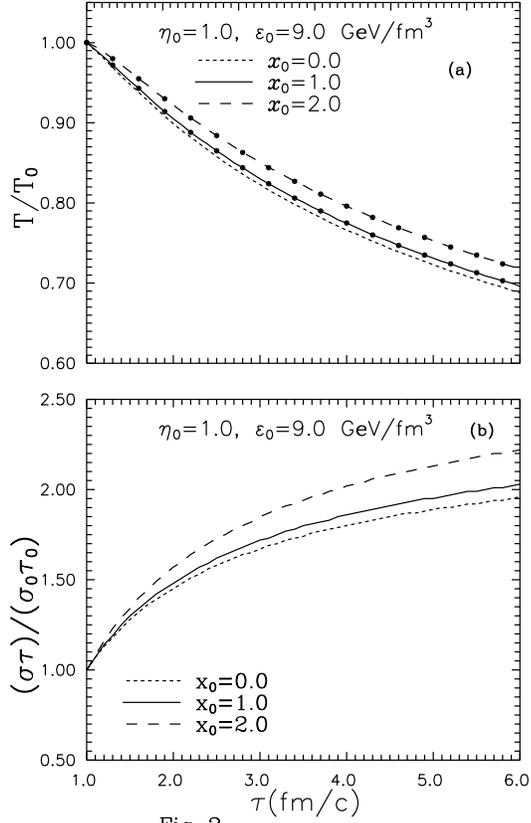
\begin{figure}[t!]
\vspace{3cm}
\hspace{3cm}
\begin{minipage}[t]{12cm}
\psfig{figure=fig2.out,width=12cm,height=12cm}
\caption{(a) The ratio of the temperature $T/T_0$
and (b) the ratio of total entropy $(\sigma \tau)/(\sigma_0\tau_0)$  as a function of
proper time $\tau$.
The solid circles are for $\mu / \mu_0$ as discussed in the text.}
\label{Figure 2}
\end{minipage}
\end{figure}
\newpage
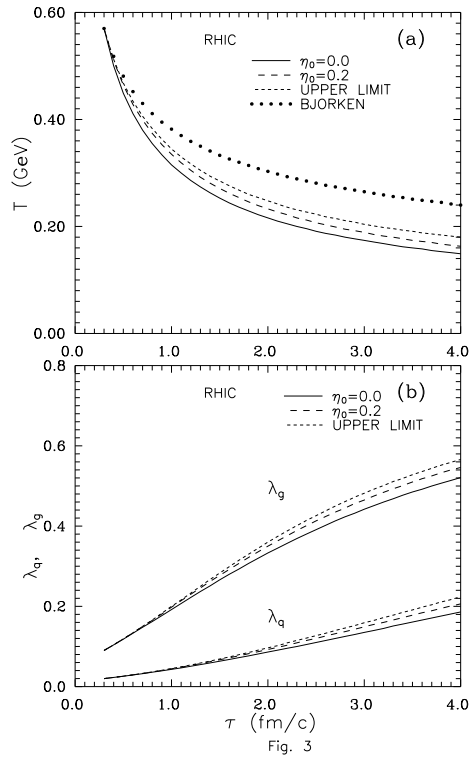
\begin{figure}[t!]
\vspace{5cm}
\hspace{3cm}
\begin{minipage}[t]{10.0cm}
\psfig{figure=fig3.out,width=10.0cm,height=10.0cm}
\caption {The evolution of (a) temperatute T , (b)gluon fugacity $\lambda_g$
and quark (anti-quark) fugacities $\lambda_q$  as a function of proper time $\tau$
with  initial conditions  $T_0$=0.57 GeV,~~ $\tau_0$=0.31 fm,
~~$\lambda_{g0}$= 0.09 and $\lambda_{q0}=\lambda_{\bar q0}$=0.02
for $\eta_0$=0.0 and 0.2. The short-dashed curves represent
the upper limit when $\eta=\frac {\epsilon \tau}{4}$. The curve with solid circles
in (a) is the Bjorken's limit.}
\label{Figure 3}
\end{minipage}
\end{figure}
\newpage
\begin{figure}[t!]
\vspace{3cm}
\hspace{3cm}
\begin{minipage}[t]{10cm}
\psfig{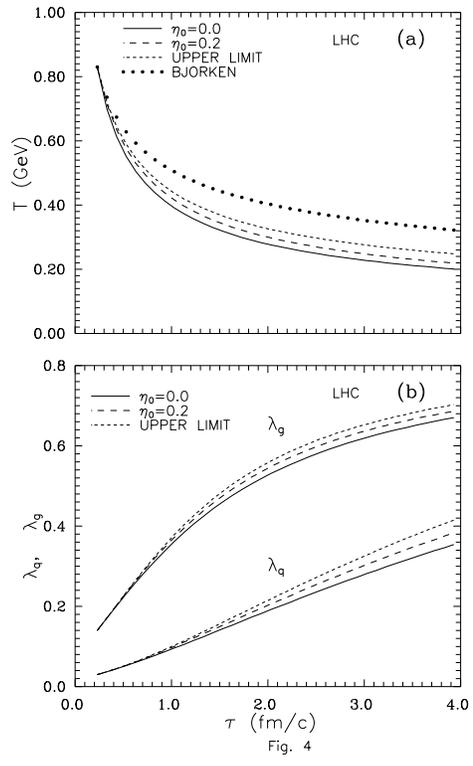}
\caption{Same as Figure 3, but for initial conditions at LHC,
i.e $T_0$=0.83 GeV, $\tau_0$=0.23 fm,
$\lambda_{g0}$= 0.14 and $\lambda_{q0}=\lambda_{\bar q0}$=0.03.}
\label{Figure 4}
\end{minipage}
\end{figure}
\end{document}